\documentclass{iau}
\usepackage{graphicx}
\newcommand{\arcsec}{\hbox{$^{\prime\prime}$}}
\newcommand{\farcs}{\mbox{$.\!\!^{\prime\prime}$}}
\title[Jet X-ray  emission] {A survey of X-ray emission from 100 kpc radio jets} 

\author[D. A. Schwartz, et al.]   
{
D. A. Schwartz$^1$
H. L. Marshall$^2$
D. M. Worrall$^3$
M. Birkinshaw$^3$\\
E. Perlman$^4$
J. E. J. Lovell$^5$
D. Jauncey$^6$
D. Murphy$^7$
J. Gelbord$^8$\\
L. Godfrey$^9$
G. Bicknell$^{10}$
}

\affiliation{
$^{1}${Harvard-Smithsonian Center for Astrophysics, 60
  Garden Street, Cambridge, MA 02138 } \\
$^{2}${Kavli Institute for Astrophysics and Space Research,
  M.I.T., 77 Massachusetts Ave.,
  Cambridge, MA 02139} \\
$^{3}${HH Wills Physics Laboratory, University of Bristol, Tyndall
Avenue, Bristol BS8 1TL, UK} \\
$^{4}${Physics and Space Sciences, Florida
  Institute of Technology, 150 West University, Melbourne,
  FL  32901} \\
$^{5}${School of Mathematics and Physics, Private Bag 21,
  University of Tasmania, Hobart Tas 7001, Australia}\\
$^{6}${CSIRO Australia Telescope National Facility, PO Box
  76, Epping NSW 1710, Australia} \\
$^{7}${Jet Propulsion Laboratory, 4800 Oak Grove Drive,
  Pasadena, CA 91109} \\
$^{8}${The Pennsylvania State University, University Park, State College, PA 16801}\\
$^{9}${ASTRON, PO Box 2, 7990 AA Dwingeloo, The Netherlands} \\
$^{10}${Research School of Astronomy and Astrophysics,
  Australian National University, Cotter Road, Weston Creek, Canberra,
ACT72611, Australia}\\
email: {das@cfa.harvard.edu}
}

\pubyear{2014}
\volume{313} 
\pagerange{xx--xx}
\jname{Extragalactic jets from every angle}
\editors{F. Massaro, C.C. Cheung, E. Lopez, A. Siemiginowska, eds.}

\begin{document}

\maketitle

\begin{abstract}
We have completed a \emph{Chandra} snapshot survey of 54 radio jets that are
extended on arcsec scales. These are associated with flat spectrum radio quasars
spanning a redshift range $z$=0.3 to 2.1. X-ray emission is detected
from the jet of approximately 60\% of the sample objects.
 We assume minimum energy and apply conditions consistent with the
original Felten-Morrison calculations in order to  estimate
the Lorentz factors and the apparent Doppler factors.  This allows
estimates of the enthalpy fluxes, which turn out to be comparable to the
radiative luminosities.

(galaxies:) quasars: general; galaxies: jets; X-rays: jets; X-rays: quasars
\end{abstract}

\firstsection 
\section{Introduction}
We have used the \emph{Chandra} X-ray observatory (\cite{Weisskopf02,Weisskopf03,Schwartz14}) to carry out a
survey for X-ray emission from the radio jets of 54 flat-spectrum
radio quasars (\cite{Marshall05,Marshall11}).  The parent sample
consists of flat-spectrum radio sources with 5 GHz flux density
greater than 1 Jy, taken from \cite{Murphy93} based on VLA
observations, or with a flux density at 2.7 GHz greater than 0.34 Jy
from ATCA observations by \cite{Lovell97}. We considered only jets
longer than 2\arcsec\ projected on the sky in order to resolve
multiple regions with the half-arcsec resolution of the \emph{Chandra}
X-ray telescope. From that parent population we selected objects for
which either we predicted an X-ray detection in a 5 ks observation, or
which had a one-sided linear radio morphology. Those two selection
criteria were motivated by the serendipitous \emph{Chandra}
observation of PKS 0637-752 (\cite{Schwartz00,Chartas00}), scaling the
predictions of X-ray flux from the X-ray to radio jet ratio for that
object, and considering that one-sided linear morphology might
indicate bulk relativistic motion. Such a relativistically beamed jet
was the basis of the interpretations of the PKS 0637 X-ray emission as
inverse Compton (IC) scattering of the cosmic microwave background (CMB)
(\cite{Tavecchio00,Celotti01}). 

The following section  presents model-independent correlations
derived directly from the data. The final section examines what we can
learn by modeling the jet X-ray emission as inverse Compton  scattering
of electrons by the cosmic microwave  background in a relativistic jet
beamed nearly at our line of sight. In both studies we consider only
the straight portion of the jet from the quasar out to a large
apparent angular bend (where typically the X-rays disappear), and
do not include the terminal hotspot or radio lobe. Since the lifetimes
of the radio jets are inferred to be relatively short, 10$^{7-8}$ years,
the regions we are sampling may be somewhat heterogeneous due to
differing ages.

\begin{figure}[t]
\begin{center}
 \includegraphics[width=0.95\columnwidth]{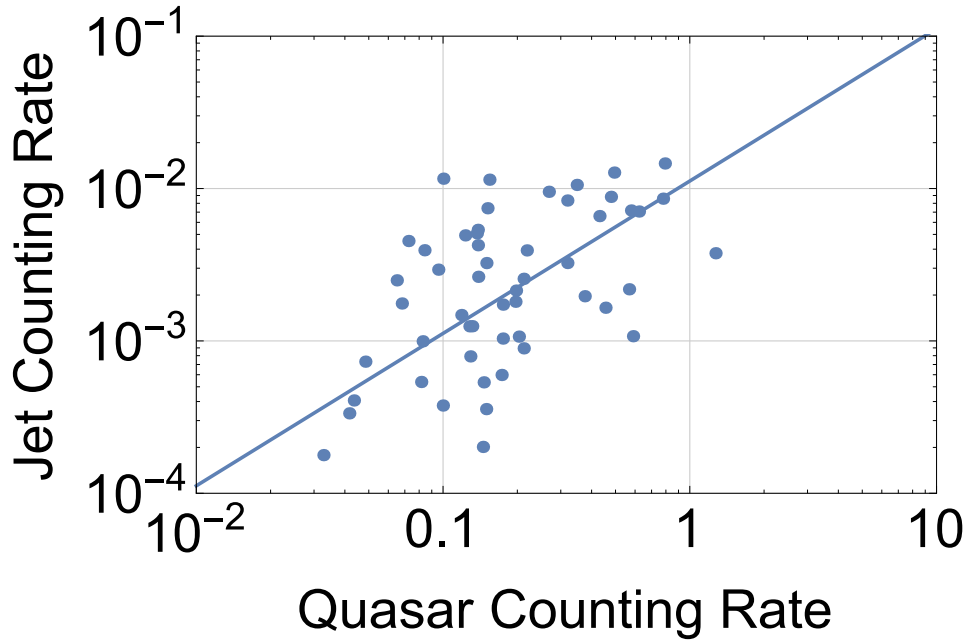}
 \caption{Raw X-ray counting rates s$^{-1}$ from the jet and from the quasar. The
   data include only the initial ``snapshot'' observation, of 5 to 10
   ks duration. Thirty-one of the 54 observations are considered a firm
   detection of X-rays from the jet. \label{fig1}}
\end{center}
\end{figure}

\section{Correlations from the data}
Figure~\ref{fig1} shows raw counting rates in the jet region vs. the
quasar core region. All observations were targeted on ACIS-S3, and
level 2 counts were taken in the 0.5 to 7 keV band. For photon power law
spectral indices in the range 1 to 2, 1 count ks$^{-1}$ is
approximately 0.94 nJy at 1 keV, or 2.2 $\times$ 10$^{-15}$ erg cm$^{-2}$
s$^{-1}$ in the 0.5 to 7 keV band. For the jet region
we used a 2\arcsec\ wide rectangle drawn on a DS9 (\cite{Joye03}) plot
of an L, C, or K band radio image. The length was extended to include
the straight part of the radio image but ended before any terminal hot
spot or lobe. The quasar core was taken to be a circle of 1\farcs26
radius. The correlation shows a large scatter, with a correlation
coefficient only 0.43, but which has a probability less than 10$^{-4}$
of no correlation. One might expect that the power in the jet is
correlated with the energy release at the black hole, hence explaining
Fig.~\ref{fig1}. However since these FR II jets are almost certainly
relativistically beamed, their apparent radiative output will vary
widely with the angle to our line of sight, and should not show a
correlation to any isotropic source of radiation.  The X-ray to
optical ratio for radio loud quasars has been known to be larger than
for radio quiet quasars and even more so for flat-spectrum radio
quasars (\cite{Worrall87}), so the present correlation may present new
evidence that the quasar core has an X-ray component that is
beamed in the direction of the kpc jet.

A simple least squares fit of the jet counting rate to the quasar
counting rate gives a mean proportionality constant 1.1\% $\pm$ 0.3\%,
to 95\% confidence, and with a scatter about a factor of 2.5.  Even if
both the jet and quasar core are beamed, the scatter would be expected
to be large since the detailed emission mechanisms, (e.g., target
photons from Compton scattering and/or contribution from synchrotron
radiation) should be different.

\begin{figure}[t]
\begin{center}
 \includegraphics[width=0.49\columnwidth]{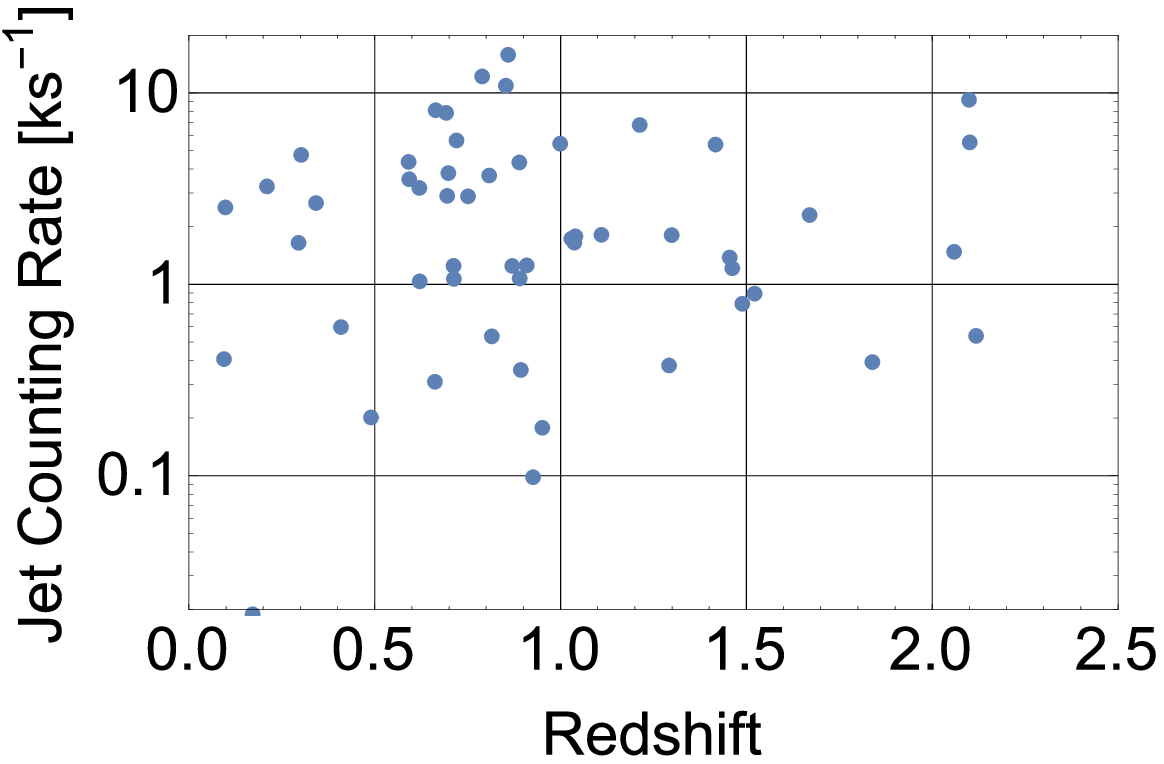}
  \includegraphics[width=0.49\columnwidth]{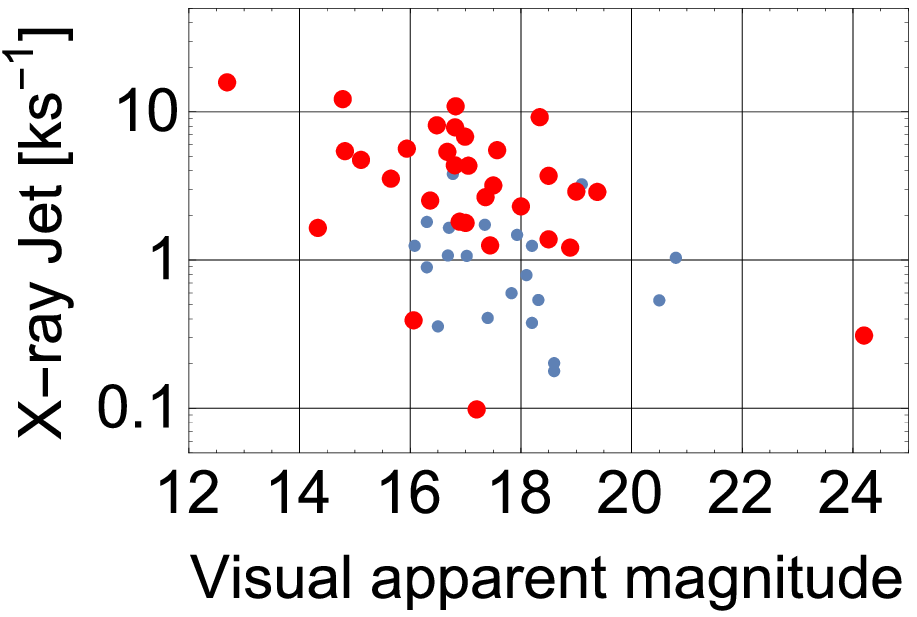}\\
 \includegraphics[width=0.49\columnwidth]{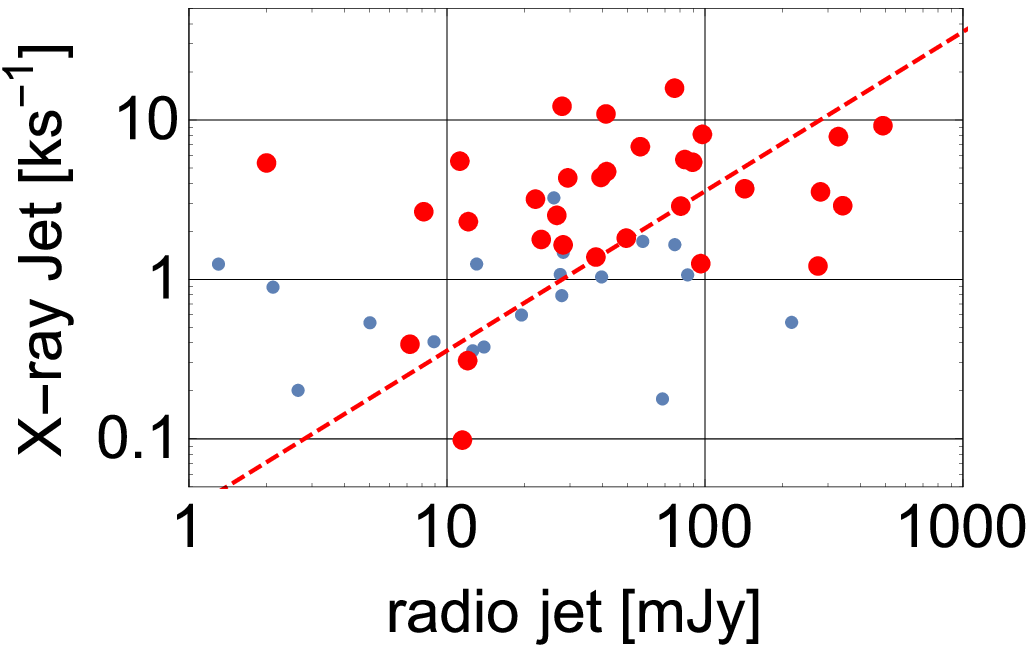}
  \includegraphics[width=0.49\columnwidth]{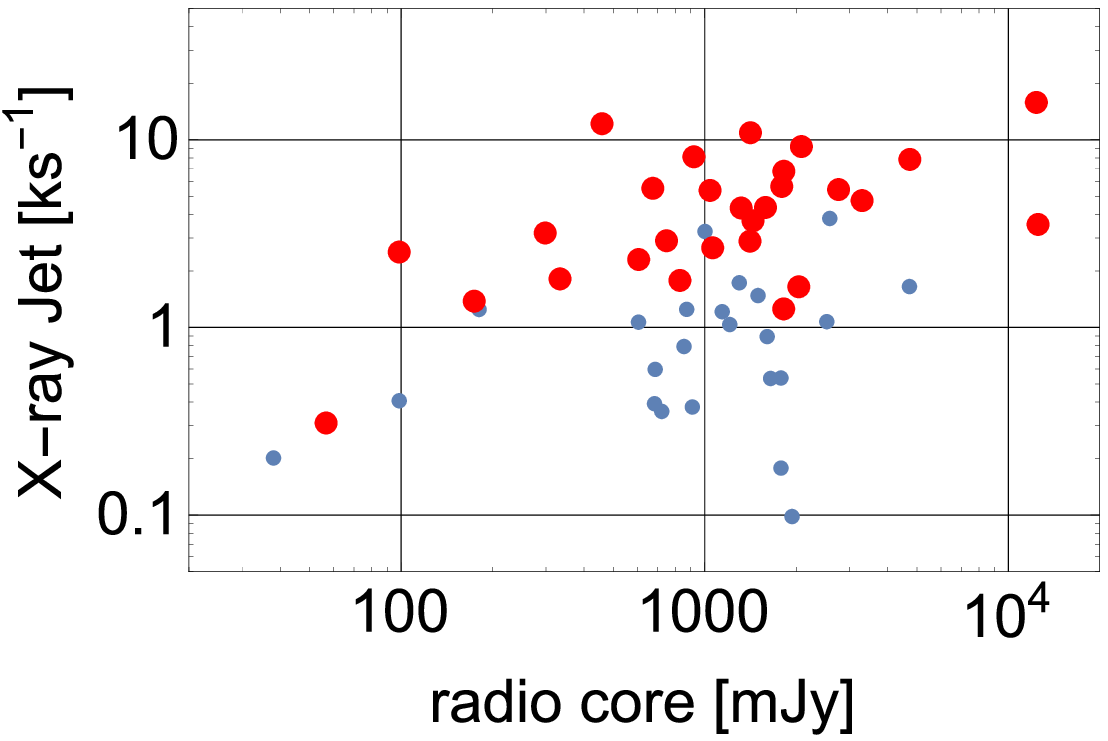}
 \caption{Correlations of the raw X-ray counting rates from the
   jets. Clockwise from upper left, the plots are against redshift,
   the quasar apparent magnitude, the radio core flux density, and the
 radio jet flux density. The larger (red in on-line version) points
 are the detected jets, the smaller (blue) points include all the
 observations. The dashed line in the lower left panel is the ratio
 observed for the jet in PKS 0637-752, and was used to select 
targets for this survey.}
   \label{fig2}
\end{center}
\end{figure}

Other relations to the X-ray flux are shown in Fig.~\ref{fig2}. There
does not appear to be any correlation with redshift or optical
magnitude. The X-rays  may correlate to the radio core flux density,
but the present figure is dominated by the half dozen points at
$\le$ 100 and $\ge$10$^4$ mJy. Correlation would be expected simply from the
fact that the quasar core X-ray flux increases with the radio
(\cite{Worrall87}). Although the predicted X-ray flux based on the
ratio of the X-ray to radio jet in PKS 0637-752, (shown as the dashed
line in the lower left panel) has served well to detect X-rays from
$\approx$ 60\% of our jet sample, the scatter is very large with no
obvious correlation.

\begin{figure}[t]
\begin{center}
 \includegraphics[angle=-90.,width=0.95\columnwidth]{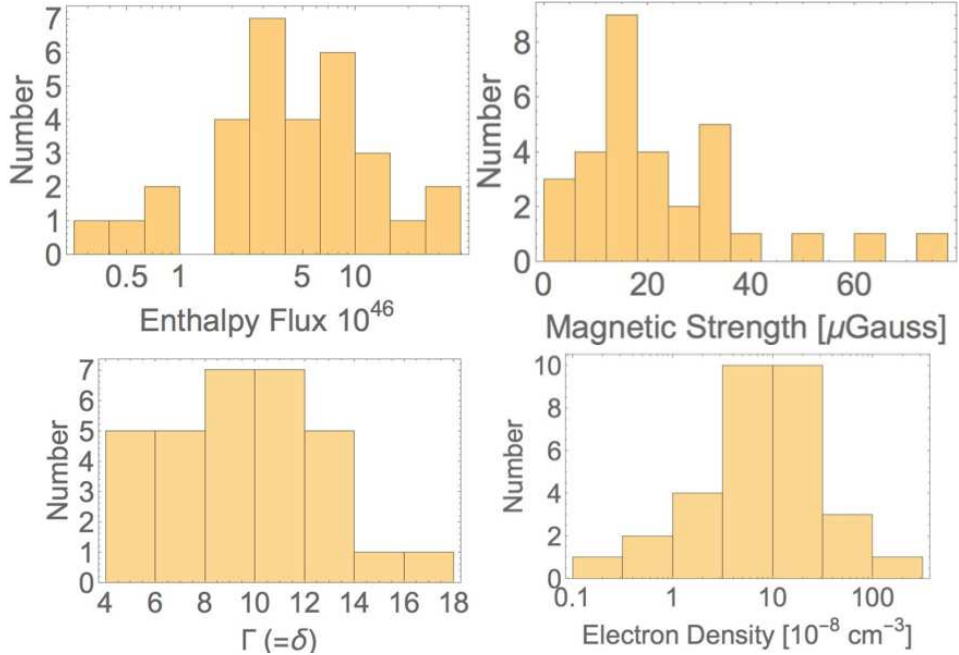} 
 \caption{Jet properties derived assuming minimum energy, IC/CMB
   emission of X-rays, Doppler factor $\delta$ equal to the bulk
   Lorentz factor $\Gamma$, and charge neutrality via equal energy
   and density of protons and electrons. See text for other details. The
   magnetic field strength and electron density are in the jet rest
   frame. The enthalpy flux [10$^{46}$ erg s$^{-1}$] and Lorentz factor are in
   the CMB frame, which is approximately the observer frame.}
   \label{fig3}
\end{center}
\end{figure}

\section{Implications of the inverse Compton/cosmic microwave background interpretation}

For almost every jet where optical data exist, the optical flux or
upper limit does not allow the jet X-ray spectrum to be an extension
of the synchrotron radio spectrum of the jet. Although exceptions
exist, this has led to the general interpretation of the X-ray
emission as inverse Compton  scattering of the cosmic microwave
background. We apply the formalism previously used by
\cite{Schwartz06}, based on work by \cite{Felten66} and
\cite{Bicknell94}. Briefly, we assume minimum-energy conditions
producing the radio synchrotron emission, with a low energy cutoff to
the electron spectrum at Lorentz factor $\gamma_{\rm min}$=30, and with
relativistic beaming so that the energy density of the CMB is enhanced
by a factor $\Gamma^2$ in the jet rest frame. We use the supersnapshot
formalism discussed in \cite{Jester08} to calculate the rest frame
volume as V=V$_{\rm observed}/(\delta \sin\theta)$. Further assumptions
must be made to break the degeneracy between the Doppler beaming factor,
$\delta=1/(\Gamma(1-\beta \cos\theta))$, the bulk Lorentz factor of
the jet, $\Gamma$, and the angle $\theta$ to the observer's line of
sight.  The electron spectrum producing the radio synchrotron is
assumed to extend to sufficiently low energies to produce the X-rays.
Roughly, $\gamma \approx$ 1000/$\delta$ produces 1 keV X-rays. For PKS
0637-752, \cite{Mueller09} show that $\gamma \le$ 70 to produce the
lowest energy X-rays; consistent with our choice of $\gamma_{\rm min}$=30. We
assume homogeneity and isotropy for the particles and magnetic fields
in the jet rest frame.

For an initial evaluation of the entire sample, we have taken
$\delta$=$\Gamma$. This places the jet at the largest allowed angle for the
given value of $\delta$, $\sin \theta$=1/$\delta$. For this case, the photons
seen by the observer are emitted in the jet frame perpendicular to the
direction of propagation of the jet, and V=V$_{\rm observed}$. It also implies
$\delta$ is half its maximum possible value of 2$\Gamma$, which is approached
for large $\Gamma$ and small $\theta$. Fig.~\ref{fig3} summarizes the results
for the 31 detected jets. For these values we have also assumed equal kinetic
energy in relativistic protons and electrons, and equal number density of
protons and electrons, which can both be satisfied since we have no constraints
on the proton spectral shape. We can summarize the sample properties as $\Gamma
\approx$ 10, magnetic field strength $\approx$ 10--20 $\mu$G (1--2 nTesla),
electron density $\approx$ 10$^{-7}$ cm$^{-3}$ and enthalpy flux $\approx$
(5--10) $\times$ 10$^{46}$ erg s$^{-1}$.

\begin{figure}[t]
\begin{center}
 \includegraphics[angle=-90.,width=0.95\columnwidth]{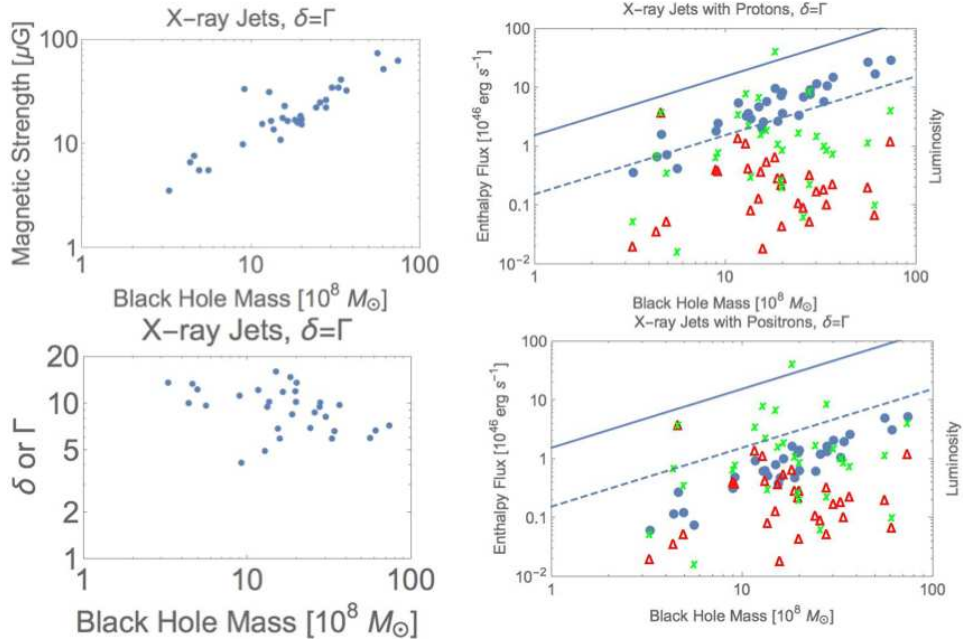}
 \caption{Derived properties of the 31 detected jets, plotted against a proxy
   for black hole mass. Clockwise from top left: Magnetic field strength,
   Enthalpy flux assuming proton density equals electron density, Enthalpy flux
   assuming proton density is zero, Bulk Lorentz factor of the jet. In each
   plot, the small (blue in on-line version) dots are the jet X-ray data. In
   the enthalpy plots, the (green) crosses are the optical luminosities of the
   quasar and the (red) triangles are the X-ray luminosities. }
   \label{fig4}
\end{center}
\end{figure}

Fig.~\ref{fig4} plots the derived quantities vs. a proxy for black hole
mass. The mass estimate is taken from the relation in \cite{Gultekin09}:
M$_{\odot}$=10$^{0.19}$L$_{\rm r 38}^{0.48}$L$_{\rm x 40}^{-0.24}$, where
L$_{\rm r 38}$ is the radio luminosity of the quasar in units of 10$^{38}$ erg
s$^{-1}$ and L$_{\rm x 40}$ is the quasar X-ray luminosity in units of
10$^{40}$erg s$^{-1}$. The magnetic field strength, electron density (which is
forced to be proportional to the magnetic field energy density by the minimum
energy assumption), and enthalpy flux all correlate to the black hole mass
proxy. This is due at least in part to the correlation of the jet and quasar
X-ray flux, since the latter enters the equation for the mass proxy.

Assuming positrons, rather than protons, provide charge neutrality in
the jet lowers the estimates of the enthalpy flux by a factor of about
5. In either case the energy flux carried by the jet is comparable to
the radiative luminosity, or at least a significant component of the
accretion energy budget. Other derived quantities depend only slightly
on the positive charge carrier, typically decreasing about 20\% from
all protons (e.g., Fig.~\ref{fig3}) to all positrons.

\vspace{4mm}

This research has made use of SAOImage DS9, developed by Smithsonian
Astrophysical Observatory, of NASA's Astrophysics Data System, and of
the NASA/IPAC Extragalactic Database (NED) which is operated by the
Jet Propulsion Laboratory, California Institute of Technology, under
contract with the National Aeronautics and Space Administration.
Support for this work was provided in part by the National Aeronautics
and Space Administration (NASA) through the Smithsonian Astrophysical
Observatory (SAO) contract SV3-73016 to MIT for support of the Chandra
X-Ray Center (CXC), which is operated by SAO for and on behalf of NASA
under contract NAS8-03060. Support was also provided by NASA under
contract NAS 8-39073 to SAO, and by grants GO9-0121B from the CXC and
GO-11838.04-A from the Space Telescope Science Institute.

\end{document}